\documentclass[twocolumn,amssymb,pre]{revtex4}
\usepackage{graphicx}  
\usepackage{graphics}
\usepackage{epsfig}
\usepackage{amsmath}
\usepackage{gensymb}
\usepackage{color,hyperref}
\newcommand{\AW}[1]{{\color{black}#1}}

\begin{document}
\title{Cost and benefits of CRISPR spacer acquisition}
\author{Serena Bradde$^{1}$, Thierry Mora$^{2*}$, Aleksandra M. Walczak$^{3*}$}
\address{	$^{1}$American Physical Society, Ridge, NY, USA  
and David Rittenhouse Laboratories, University of Pennsylvania, Philadelphia, PA 19104 \\
$^{2}$ Laboratoire de physique statistique, CNRS, Sorbonne Universit\'e, Universit\'e Paris-Diderot,
          and \'Ecole Normale Sup\'erieure (PSL University), 24, rue Lhomond, 75005 Paris, France\\
$^{3}$Laboratoire de physique th\'eorique, CNRS, Sorbonne
Universit\'e, and \'Ecole Normale Sup\'erieure (PSL University), 24,
rue Lhomond, 75005 Paris, France\\
$^*$ These authors contributed equally.
}

\begin{abstract}
CRISPR-Cas mediated immunity in bacteria allows bacterial populations to protect themselves against pathogens. However, it also exposes them to the dangers of auto-immunity by developing protection that targets its own genome. Using a simple model of the coupled dynamics of phage and bacterial populations, we explore \AW{how acquisition rates affect the survival rate of the bacterial colony}. We find that the optimal strategy depends on the initial population sizes of both viruses and bacteria. Additionally, certain combinations of acquisition and dynamical rates and initial population sizes guarantee protection, due to a dynamical balance between the evolving population sizes, without relying on acquisition of viral spacers. Outside this regime, the high cost of auto-immunity limits the acquisition rate. We discuss these optimal survival strategies in terms of recent experiments. 
\end{abstract}
\maketitle

\section{Introduction}

Organisms have developed a wide variety of strategies to deal with pathogens \cite{Medzhitov1997, Boehm2011, Spoel2012, Murphy2007,Mayer2015b}. Bacteria use immunity both at the population level by adopting heterogenous phenotypes with different susceptibility to varying environments \cite{Levins1968,Seger1987,Donaldson-Matasci2008,Mayer2017}, as well as specific solutions that target invading phage viruses \cite{Abedon2012, Barrangou2014, Labrie2010}. These strategies involve restriction enzymes that render the viral genetic material unviable \cite{Abedon2012}, and the CRISPR (Clustered Regularly Interspaced Short Palindromic Repeats) Cas (CRISPR-associated system of proteins) system \cite{Barrangou2014, Labrie2010}, which allows bacterial lineages to develop memory about encountered pathogens and protect future generations. 

\begin{figure}
 	 \includegraphics[width=\linewidth]{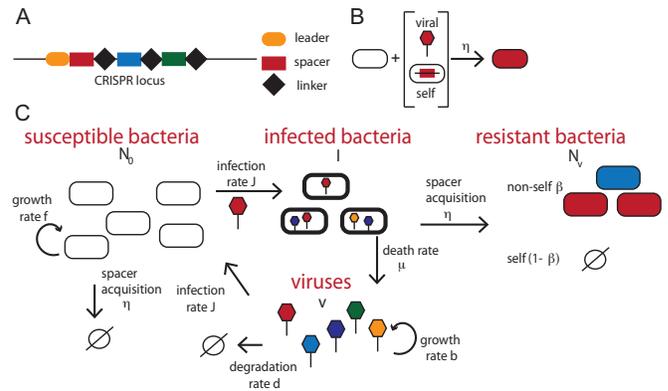}
	\caption{\label{fig:intro}\textbf{Cartoon representation of the CRISPR-Cas system model}. A. Bacteria protect themselves from surrounding phage viruses, $V$, by up-taking segments of their DNA and incorporating it into their genome. The up-taken spacers (colored rectangles) are incorporated into the CRISPR locus and separated by linkers (black diamonds). B. A susceptible bacterium (empty oval) can take up viral DNA spacers from the red phage (top) or its own DNA as spacers from its own genome (bottom). The acquisition rate of any new spacer is $\eta$. \AW{We note that we define our rates as the proportionality constant to the number of individuals in the reacting populations, as is customary in chemical kinetics. Our rates have units of inverse time.} C. In a given viral environment, a number $N$ of bacteria do not carry protection and are phage susceptible (empty ovals), they can get infected (empty ovals with viruses) and after acquisition bacteria become resistant with the appropriate anti-phage spacers (colored ovals). \AW{All bacteria reproduce at a maximum growth rate $f$ up to carrying capacity $K$. Susceptible bacteria become infected with rate $J$ or they acquire a self-targeting spacer with rate $\eta$ (since they are not infected they can only acquire self-targetting spacers). Infected cells can acquire a spacer with rate $\eta$ and become resistant with probability $\beta$ or die due to autoimmunity from self-spacers. Infected cells that do not acquire spacers die at a rate $\mu$ producing $B$ new virus copies.} \AW{We assume there are two type of spacers: the self-targeting spacers of bacterial origin, which lead to the death of the bacteria through auto-immunity, and a proportion  of phage-targeting spacers up-taken with probability $\beta$, whose incorporation turns susceptible bacteria into protected ones.}}
\end{figure}

The CRISPR-Cas mechanism consists of a machinery of enzymes that integrates $20$ to $50$ base pair (bp) unique viral DNA or RNA fragments, \AW{also called {\it spacers}, into the CRISPR loci of the bacterial genome} (see Figure~\ref{fig:intro}A). Upon expression, \AW{these spacer RNAs serve as a guide to identify viral genomic material} and target it for degradation. Through its ability to constantly integrate new genetic material, CRISPR-Cas provides bacteria with adaptive immunity, which is heritable by offspring thanks to its integration into the host genome \cite{Barrangou2014, Labrie2010}. Additionally to acquisition of new spacers, spacers can also be lost. In fact, the distribution of diversity and abundance of individual spacers in virally challenged bacterial populations is highly variable \cite{Bonsma-Fisher2018, Heidelberg2009, Andersson2008, Tyson2008, Paez-Espino2015}. Acquired spacers in type I and II systems, which are the focus of this paper, are not uniformly sampled from the viral genomes but are chosen for excision by Cas proteins from fragments adjacent to short (3-5 bp) regions called protospacer adjacent motifs (PAM) \cite{Mojica2009}. While mechanisms for avoiding targeting the CRISPR-Cas \AW{loci} itself have been identified \cite{Marraffini2010a}, since PAM sequences are so short they can also \AW{be found in other parts of} the bacterial host's own DNA and acquire self-targeting spacers \cite{Levy2015b, Wei2015} (see Figure~\ref{fig:intro}B). Additionally, self-targeting spacer related mortality has been proposed as one of the major costs associated with the CRISPR-Cas system \cite{Westra2015, Stern2010, Vercoe2013,Edgar2010}.

A large body of theoretical work has focused on the evolution of the spacer cassettes~\cite{Berezovskaya2014, Childs2012, Childs2014, Han2017} and their diversity~\cite{He2010,Haerter2011, Weinberger2012a, Bradde2017, Bonsma-Fisher2018}. These studies range from exploring spacer distribution dynamics as a function of the parameters of the system such as the rates of spacer acquisition and loss and mutation rates, to more detailed explanations of the spatial structure of spacer insertions. Apart from its Lamarckian evolution aspect \cite{Koonin2009,Haerter2012}, a lot of models have studied the CRISPR-Cas system from the point of view of bacterial-viral co-evolution and the role of the ecological context \cite{Iranzo2013, Levin2013, Westra2015, PleskaNatEE2018}, with an emphasis on the cost of maintaining the CRISPR-Cas system \cite{Levin2010, Westra2015,Vale2015a}. \AW{Studies have also focussed on other bacterial defence strategies that play an important role, such as restriction modification enzymes \cite{Labrie2010, Pleska2016}. } More recently, aspects of CRISPR dynamics have been studied \AW{that involve interactions between members of the population, including} communication between individual bacteria via quorum sensing mechanisms \cite{Bonsma-Fisher2018}.

The CRISPR-Cas system seems like an efficient immune strategy, with evolutionary advantages particularly for very rare pathogens \cite{Mayer2015b}. However, it carries potential metabolic costs \cite{Vale2015a}. Additionally, similarly to other specific immune systems, such as the adaptive immune system of vertebrates \cite{Perelson1997}, it carries the potential risk of autoimmunity \cite{Stern2010,Marraffini2010a}. Specifically, in addition to integrating the DNA of attacking viruses, \AW{it is not currently clear how to prevent } the bacterium from integrating its own DNA into the CRISPR cassettes and then targeting itself. \AW{Some evidence  suggests that most of the acquisition is self-targeting when the interference part is turned off \cite{Wei2015}. This may be explained by cas9 and/or the cas complexes that drive recognition, playing a role during the acquisition process as suggested in the same article \cite{Wei2015,Datsenko2012}, or cas proteins being unable to distinguish self versus non-self targeting spacers effectively. While bacterial immunity, even in controlled experiments, is not limited to the CRISPR-Cas system, and involves the co-evolution of the bacterial and phage populations, here we focus only on the specific consequences of potentially acquiring autoimmunity via self-spacer uptake on short timescales.}  
\AW{ We ask how bacteria can survive, even in the worst case scenario when the DNA acquisition machinery is unbiased. We want to show that there is an optimal way that bacteria may regulate acquisition,} so that it provides them with the largest protection against foreign viruses, \AW{while minimizing the costs of self-targeting.}

\begin{figure*}
	 \includegraphics[width=\linewidth]{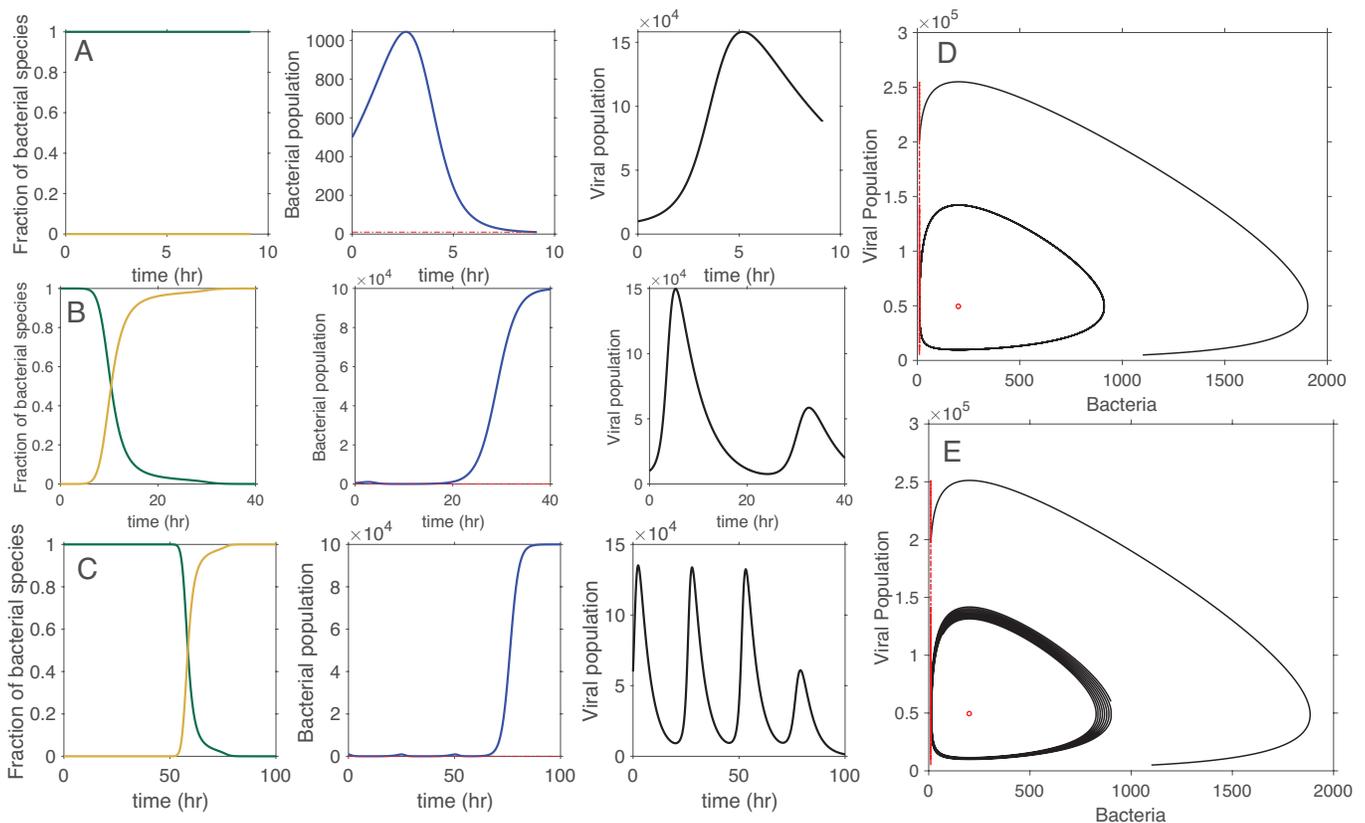} 
	{\caption{\label{fig:cartoon}\textbf{ Typical behavior of the bacterial-viral model}. The left panels show the time evolution of the fraction of bacterial species (green are the susceptible bacteria and yellow are the resistant bacteria), the number of bacteria and the number of phage with parameters $\beta=0.01$, $K=10^5$, $d=0.2\;\mbox{hr}^{-1}$ per phage and $b=10^{-3}\;\mbox{hr}^{-1}$ per phage, $\mu=0.8\;\mbox{hr}^{-1}$ per cell, in three regimes: (A) bacteria go extinct ($\eta=0.001\;\mbox{hr}^{-1}$ per cell, and initial conditions $N_0=500, V_0=\times 10^{4}$), (B) virus goes extinct and bacteria are saved by immunity ($\eta=0.02\;\mbox{hr}^{-1}$ per cell,  and initial conditions $N_0=500, V_0=\times 10^{4}$), (C) virus and bacteria co-exist oscillating and then acquisition happens at a stochastic time ($\eta=0.03\;\mbox{hr}^{-1}$ per cell and initial conditions $N_0=900, V_0= 6\times 10^{4}$). (D) Phase-space diagram showing bacterial population as a function of phage population for  $\beta=0.01$ $\eta=0.005\;\mbox{hr}^{-1}$ per cell for two initial conditions $[900,6\times 10^4]$ and $[1100, 5\times 10^3]$ in the conservative case $K\gg N_{\rm tot}$. (E) Phase-space diagram for the same parameters as in panel B in the non-conservative dynamics.}}
\end{figure*}

\section{Model}
We consider an environment with $V$ copies of the virus, and two types of bacterial cells (see Fig.~\ref{fig:intro} C): $N$ -- the number of bacterial cells without spacers conferring protection against the virus, and $N_{\rm v}$ -- the number of immune bacterial cells thanks to the presence of a specific spacer. The total number of bacteria is $N_{\rm tot}=N+N_{\rm v}$. For simplicity, we assume a single, constant viral strain, and we do not model the spacer composition in detail. Spacer-carrying bacteria survive encounters with the virus, while susceptible bacteria do not unless an acquisition event occurs.
\AW{ Both bacterial populations grow  with a maximum growth rate $f$, but are limited by resources encoded by the  carrying capacity $K$.  Susceptible cells get infected by the virus with rate $J$, \AW{infected cells die at rate $\mu$  producing $B$ copies of the virus}. Infected cells can also acquire spacers with rate $\eta$. The fraction of spacers that are acquired from exogenous genetic material is encoded by the parameter, $\beta$, which describes the probability of non-self spacer acquisition, so that the rate of exogenous spacer acquisition is $\beta\eta$.  Bacteria also acquire self-targeting spacers with an overall rate $(1-\beta)\eta$. Within this model, the mean growth rate of viral cells is given by bursting of infected cells that do not acquire spacers. Viral cells decay with rate $d$. }
At the time of infection, $t=0$, the two viral and susceptible bacterial populations are mixed together, with variable initial sizes, $N=N_0$, $V=V_0$, and $N_{\rm v}=0$, and the system evolves towards co-existence or the extinction of one or more of the species involved.

\AW{The non-self spacer acquisition parameter $\beta$ is assumed to be}  small and set by constraints beyond bacteria's control, such as the ratio of viral exogenous versus self genetic material. \AW{In principle this assumption can be relaxed and $\beta$ could depend on the multiplicity of infection (MOI) -- a quantity related to the ratio of bacterial versus viral genetic material in the cell. In this case, the non-self and self-acquisition parameters are subject to a feedback mechanism described by sigmoid functions.  For simplicity we  assume that the infection starts at low MOI so that $\beta$ can be considered constant, which is a limiting case of the fuller feedback model. On the other hand, we assume that the acquisition rate of spacers, $\eta$, is regulated  by the bacterial host population by phenotypic adaptation or quorum sensing \cite{Patterson2016, Hoyland-Kroghsbo2017}. We then ask how a bacterial acquisition rate affects the long-term probability of bacterial survival.}

The set of reactions presented above and summarized in Fig.~\ref{fig:intro} is the basis for a stochastic dynamical process, in which the numbers of viruses, susceptible, infected and non-susceptible bacteria evolve over time. An example trace for the temporal change in the numbers of the participating species shows that both bacteria and viruses can go extinct. Starting from high initial viral numbers and no immune bacteria, the number of susceptible bacteria decreases significantly, going extinct for some parameter values (Fig.~\ref{fig:cartoon}A). However, if bacteria avoid extinction, the number of immune bacteria increases and the viruses can lose their reproductive reservoir and go extinct (Fig.~\ref{fig:cartoon}B). In a third parameter regime, susceptible bacteria and viruses can co-exist in stable oscillatory dynamics (Fig.~\ref{fig:cartoon}C) until immunity is acquired and spreads throughout the bacterial population.

\AW{The process described above, before any viral spacer is incorportated, is encoded in a set of ordinary differential equations for the temporal evolution of the susceptible and infected bacterial populations, and the viral population
\begin{equation}\label{starteqns1}
\begin{split}
&\frac{\partial N}{\partial t} = \left(1-\frac{N}{K}\right)f N -  \eta N -  J V N\\
&\frac{\partial I}{\partial t}  = JVN - \eta I - \mu I,\\
&\frac{\partial V}{\partial t}= B \mu I - d V,
\end{split}
\end{equation}
where $N$ is the number of cells in the susceptible bacterial population,  $I$ is the number of cells in the infected bacterial population, and $V$ is the number of cells in the phage population. Bacteria grow with  a maximum growth rate $f$, limited by the carrying capacity $K$. Bacteria \AW{get infected} with rate $J$ and \AW{die due to} self-spacer acquisition with rate $\eta$ (since we assumed no viral spacer acquisition at this stage). The number of infected cells \AW{that are not acquiring viral targeting spacers} increases due to viral infection and decreases with rate $\eta$ due to acquisition and $\mu$ due to cell bursting. \AW{The number of viruses} grows due to bursting of infected cells \AW{producing $B$ new copies of the virus} and decays with rate $d$.  If we assume that the time scale for bursting is \AW{much smaller} than the timescale of spacer acquisition, infected cells are in steady state, $I \approx JVN/ (\eta + \mu)$. In this limit,  equations~\ref{starteqns1} simplify to 
\begin{equation}\label{starteqns}
\begin{split}
\frac{1}{N}\frac{\partial N}{\partial t}& = \left(1-\frac{N}{K}\right)f-  \eta - J V,\\
\frac{1}{V}\frac{\partial V}{\partial t} &= (1 - \eta/\mu){B} J N- d,
\end{split}
\end{equation}
} 


Rescaling $V$ by $J$, $v=JV$ these equations simplify to:
\begin{equation}\label{rescaledeqns}
\begin{split}
\frac{1}{N}\frac{\partial N}{\partial t}& = \left(1-\frac{N}{K}\right)f-\eta - v,\\
\frac{1}{v}\frac{\partial v}{\partial t} &= b N (1 - \eta/\mu)- d,
\end{split}
\end{equation}
\AW{where $b=BJ$ is the rescaled bursting factor. }

These equations (Eqs.~\ref{starteqns} and~\ref{rescaledeqns}) are valid as long as protected bacteria are completely absent, $N_{\rm v}=0$. When acquisition, which is a random event, first occurs, the bacterial population as a whole is likely to survive over very long time scales, as the protected population will displace the susceptible one and eventually drive the viral population to extinction. 
This scenario, in which all three species are coupled, is not described by Eq.~\ref{starteqns}, which only models the pre-acquisition period. Thus, while $N$ and $V$ are treated deterministically, $N_{\rm v}$ is treated stochastically through the first time of acquisition, {\em i.e.} the time at which $N_{\rm v}$ jumps from 0 to 1. At that point, we consider the bacterial population safe. Clearly this is an approximation, which we discuss later.

The first acquisition event follows a Poisson point process occurring with rate \AW{$(\beta/\mu)  \eta J VN$}, meaning that it occurs with probability \AW{$(\beta/\mu)\eta J VN dt$} between times $t$ and $t+dt$. Thus the probability that no acquisition event occurs before time $t$ is:
\begin{equation}\label{ac_prob}
\mathcal{P}(t)=e^{-(\beta /\mu) \eta J\int_0^{t} N(t) V(t) dt}=e^{-(\beta/\mu)\eta\int_0^{t} N(t) v(t) dt}.
\end{equation}
Before that event, the bacterial population may be driven to \AW{very low concentrations by the virus, putting the full colony at risk.} We assume that this happens when \AW{the susceptible bacteria numbers} $N$ reach some low population size threshold $N_{\rm ext}$. \AW{ Within this model we are not considering any other form of immunity.} This extinction boundary is chosen in an arbitrary way to reflect the small size of a population near extinction and to avoid explicitly treating the stochastic dynamics at small copy numbers. \AW{By the term `avoiding extinction,' we mean the possibility that susceptible bacteria survive the small population size bottleneck just by chance. In most of the laboratory experiments, bacteria never go extinct because they mutate and become resistant by other mechanisms (restriction-modification mechanisms (RM) or surface modification mutants (SM)) that we do not consider. Here we do not compare different types of immunity but address the question of whether the CRISPR-Cas can be beneficial, despite self-targeting.} 
The time $t_{\rm ext}$ at which extinction occurs is determined by the solution to the deterministic equation Eq.~\ref{rescaledeqns} ($t_{\rm ext}$ can take a value of $+\infty$ if the threshold is never reached, see below). In this formulation and within these approximations, the bacterial population will survive if and only if an acquisition event occurs before extinction, with probability $P_{\rm surv}=1-P_{\rm ext}=1-\mathcal{P}(t_{\rm ext})$. In the special case where no extinction occurs even in the absence of any acquisition, $t_{\rm ext}=\infty$, we obtain $P_{\rm surv}=1$.

To gain some intuition, we first describe the pre-acquisition dynamics in the limit of small bacterial sizes, $N\ll K$, where Eq.~\ref{rescaledeqns} can be integrated analytically. In that limit, the term $N/K$ drops from Eq.~\ref{rescaledeqns}, \AW{which reduces to the well-studied Lokta-Volterra equations \cite{Lotka20, Volterra56, Levin77, Weitz:2015aa}. This }approximation is justified if we are interested in small, vulnerable bacterial populations which are far from the limits imposed by their carrying capacity. \AW{As is already known \cite{Lotka20, Volterra56,  Weitz:2015aa}}, Lotka--Volterra equations admit a conserved quantity, which is constant over time:
\begin{eqnarray}
L=(f-\eta) \ln v(t) +d \ln N(t) - v(t) -b (1-\eta/\mu) N(t).\nonumber\\
\end{eqnarray}
This conserved quantity implies that the dynamics in the $(N,v)$ phase space follows a closed counterclockwise orbit around the fixed point at $v^*=f-\eta$ and $N^*=d/b(1-\eta/\mu)$ (Fig.~\ref{fig:cartoon}D). The orbit is set by the values of the initial viral and susceptible populations and the model parameters, which determine $L$ once and for all.
Small orbits that stay close to the fixed point will not hit the extinction boundary $N_{\rm ext}$. The set of initial conditions falling into these orbits defines a `safe zone', in which the bacterial population is guaranteed to survive. The critical value $L_0$ of the conserved quantity separating safe from extinction orbits is obtained by considering the critical orbit hitting the extinction boundary tangentially, so that $N=N_{\rm ext}$ and $dN/dt=0$, which implies $v=v^*=f-\eta$ and thus $L_0=L(N_{\rm ext},f-\eta)$. Orbits with $L>L_0$ will be safe, while orbits with $L<L_0$ will go extinct unless an acquisition event occurs. Initial conditions that are close to the fixed point belong to safe orbits, while either very small of very large initial viral and bacterial populations may lead to extinction.

The general, non-conservative dynamics (Eq.~\ref{rescaledeqns}) with finite $K$ behave similarly, with the difference that the orbits become spirals converging counterclockwise to the fixed point (see Fig.~\ref{fig:cartoon}E). The spirals circle inwards toward the fixed point $N^*=d/b(1-\eta/\mu)$, $v^*=(1-d/ b (1-\eta/\mu)K )f-\eta$, unless they hit the extinction boundary $N_{\rm ext}$ first.
Along the way, a susceptible bacteria may acquire a protective spacer. As in the conservative case, the probability of survival depends on the initial sizes of the viral and bacterial populations in two ways. Firstly, the longer the trajectory, the larger the probability of acquiring a spacer. Secondly, certain initial conditions ensure that the trajectory never reaches the extinction threshold defining a safe zone, just as in the conservative case. However, belonging to the safe zone can no longer be associated to the value of a conserved quantity. 

This basic qualitative analysis points to the importance of the initial conditions for the probability of survival. While this may seem as a problematic artefact of the model, recent experiments have reported that bacterial survival rates in controlled environments depend on the initial concentration ratios of phages and bacteria \cite{Westra2015}. This observation was driven by theoretical work that explored the limiting role of environmental resources on the choice between inducible and constitutive mechanisms \cite{Shudo2001, Hamilton2008}. 
Specifically, WT {\it Pseudomas aeruginosa} bacteria grown in $10^4$, $10^7$ and $10^9$  plaque forming units (pfu) of non-targeted phage (NT$\phi$) showed a large decrease in the fraction of bacteria that acquired CRISPR-Cas mediated immunity in populations with large phage exposure  (Fig. 3A in Ref.~\cite{Westra2015}). Keeping the same number of infecting phages but increasing the \AW{flask} size to decrease the phage density resulted in a higher fraction of bacteria that acquired CRISPR-Cas mediated immunity in populations with low phage density exposure (Fig. 3B in Ref.~\cite{Westra2015}). In the same experiments, opposite trends were reported for constitutive immunity quantified in terms of the loss of the phage receptor (surface modification mediated resistance). These results were linked to the costs of the two different protection modes. \AW{We have not included an alternative mechanism of defence so we can not directly compare our results to these experiments. However, these experimental results suggest that the \AW{likelihood} of using CRISPR-Cas mediated immunity show a dependence on the initial phage exposure concentrations. Additionally, a new series of experiments \cite{Patterson2016, Hoyland-Kroghsbo2017} show a dependence of cas-expression levels on cell density, suggesting that bacteria might be able to effectively regulate their acquisition rate.}

\section{Results}
\begin{figure}
	 \includegraphics[width=\linewidth]{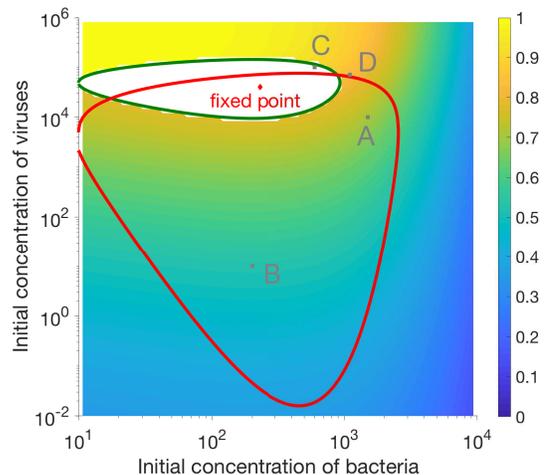}
	{\caption{\label{Fig3}\textbf{Probability of extinction for a fixed value of the acquisition rate $\eta=0.1\;\mbox{hr}^{-1}$, as function of the initial population sizes.}. The phase diagram represents the probability of extinction calculated using the coupled deterministic dynamics of Eq.~\ref{starteqns}, and Eq.~\ref{ac_prob}. The probability of extinction is defined as the probability of not acquiring a protection before reaching the extinction boundary $N=N_{\rm ext}=10$. The white area is the set of initial conditions that define the safe zone: dynamics starting from these points will never hit the extinction boundary and instead spiral counterclockwise into the fixed point. The gray diamonds A-D mark particular examples of initial conditions discussed in detail in Fig.~\ref{Fig4}. The orange line marks the safe zone for another value of $\eta=0.45\;\mbox{hr}^{-1}$. Parameters of the dynamics:  $f=0.5\; \mbox{hr}^{-1}$ per cell, $K=10^6$, $\beta=0.01$, $J=10^{-5}\;\mbox{hr}^{-1}$ per cell per phage, $\eta=0.1\;\mbox{hr}^{-1}$ per cell, $B=100$, $\mu=0.8\; \mbox{hr}^{-1}$ per phage and $d=0.2\; \mbox{hr}^{-1}$  per phage. The concentrations are given in terms of mean numbers of individuals.}}
\end{figure}

We first analyze the dependence on initial population sizes. For a fixed small value of the spacer acquisition rate $\eta=0.1\; \mbox{hr}^{-1}$ 
 and fixed growth and degradation rates, we explore the probability of survival as a function of the initial phage exposure and the initial size of the susceptible bacteria population \cite{Heler2017}. 
  We calculate the probability of extinction, defined as $P_{\rm ext}=\mathcal{P}(t_{\rm ext})$ using Eq.~\ref{ac_prob}. The results are shown in Fig.~\ref{Fig3} as a function of the initial population sizes, $N_0$ and $V_0$.

As predicted from the phase space argument of Fig.~\ref{fig:cartoon}E, a safe zone around the fixed point emerges for initial conditions that ensures that the trajectory never hits the extinction boundary (in white).
The boundary of the safe zone (black line) can be calculated by integrating the dynamical equations (Eq.~\ref{rescaledeqns}) backwards from the point of the critical trajectory that is tangent to the extinction boundary, $N=N_{\rm ext}$, and $dN/dt=0$, i.e. $v=(1-N_{\rm ext}/K)f-\eta$.

Within the safe zone, survival is ensured because the trajectory never reaches extinction, and instead reaches the fixed point (red point) even in absence of acquisition, giving the bacterial population unlimited time to uptake a protective spacer.
The safe zone is centered around the stable fixed point with a larger spread in initial bacterial population sizes (two decades for the chosen parameters) than  initial phage population sizes (one decade  for the chosen parameters).

Outside the safe zone, very large initial viral population sizes lead to rapid extinction of the bacterial population, with little chance of acquiring a spacer (yellow region). Perhaps less intuitively, very large initial bacterial populations strongly facilitate initial phage growth, leading to high-amplitude dynamics and host extinction in absence of acquisition. This effect sets the rightmost border of the safe zone. In that regime however, there is enough time and opportunity to allow for the acquisition of protection, as can be seen from the low extinction probabilities.
Likewise, small initial phage populations lead to dynamics outside the safe zone, but with a sufficient time for spacer acquisition, as the dynamics spiral counterclockwise around the fixed point (Fig.~\ref{fig:cartoon}E). As a result, a broad region of initial conditions under the safe zone ensures near-certain survival (blue region).

Crucially, the safe zone and extinction probabilities depend on the value of the acquisition rate, $\eta$. In Fig.~\ref{Fig3} we plotted, on top the results for $\eta=0.1\;\mbox{hr}^{-1}$, the boundary of the safe zone for a different value of the acquisition rate, $\eta=0.45\;\mbox{hr}^{-1}$, to illustrate how that zone can dramatically vary.
Changing $\beta$, however, does not change the boundaries of the safe zone but simply rescales the logarithm of the extinction probability by a constant factor (Eq.~\ref{ac_prob}), only affecting the values of the contours in Fig.~\ref{Fig3}.
 
 \begin{figure*}
	 \includegraphics[width=\linewidth]{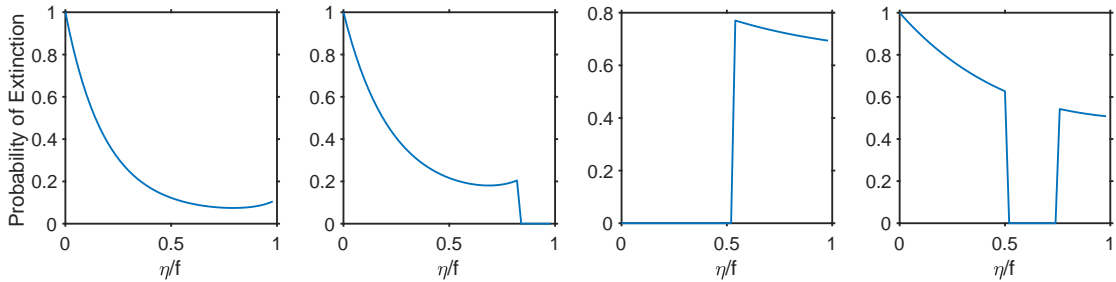}
	{\caption{\label{Fig4}\textbf{The probability of extinction as a function of $\eta$ shows 4 distinct regimes.} Panels A-D correspond to the different values of the initial conditions denoted by gray points in Fig.~\ref{Fig3}. A) For initial conditions outside the safe zone for all $\eta$, the dynamics always hit the extinction boundary regardless of the value of the aquisition rate. The extinction probability is minimized at intermediate values of $\eta$ realizing a trade-off between the cost and benefit of acquisition. B) The initial conditions are safe (i.e. lead to dynamics reaching the fixed point without ever hitting the extinction boundary) for large $\eta$ values; C) The initial conditions are safe for small $\eta$ values; D) The initial conditions are safe for intermediate $\eta$ values. Parameters of the dynamics: $K=10^6,  J=10^{-5}\;\mbox{hr}^{-1}$ per cell per phage, $f=0.5\;\mbox{hr}^{-1}$ per cell, $d=0.2\;\mbox{hr}^{-1}$ per phage, $\beta=0.001, \mu=0.8\;\mbox{hr}^{-1}$ per cell. Initial conditions (from A to D): $N_0=(1500,200,600,1100)$, $V_0=(10^4,10,10^5,7\cdot 10^4)$. }}
\end{figure*}
 
Despite increasing (on a logarithmic scale) the size of the safe zone as well as the probability to acquire protection, very large $\eta$ values are not always the optimal strategy for survival.  Depending on the initial bacterial and viral population sizes, the optimal acquisition rate that maximizes the survival probability varies dramatically (Fig.~\ref{Fig4}). We calculated the survival probability as a function of the acquisition rate $\eta$ (scaled by the growth rate $f$) for four choices of initial conditions marked by gray diamonds in Fig.~\ref{Fig3}. Since $\eta$ is an effective degradation rate for the susceptible bacterial population, $\eta$ values larger than the maximal growth rate $f$ lead to population collapse, limiting $\eta/f$ between $0$ and $1$.

For initial conditions that position the two populations away from the safe zone for all values of $\eta$, putting the susceptible population on an extinction course (point A in Fig.~\ref{Fig3}), an intermediate value of $\eta$ is optimal for bacterial survival (Fig.~\ref{Fig4}A). In this regime survival crucially depends on acquiring a protective spacer from the virus. Too small acquisition rates make it unlikely that any spacer is taken up during the spiral integration period. Conversely, too large acquisition rates increase the probability of up-taking a self-targeting spacer, which in our model leads to bacterial death and thus to a faster extinction. This tradeoff results in an intermediate value of $\eta$.

For small initial viral population sizes and small initial bacterial population sizes  (point B in Fig.~\ref{Fig3}), the dynamics fall into the safe zone for large acquisition rates (Fig.~\ref{Fig4}B). In this case, although there is still an optimal acquisition rate outside the safe zone, the optimal strategy is not to rely on luck but to choose a very large acquisition rate that will stabilize the dynamics towards the fixed point, ensuring population survival.

Note that this regime corresponds to small initial viral population sizes, since the safe zone encompasses these initial conditions only for large $\eta$: in general, the safe zone is centered around the fixed point and $v^*=f-\eta$, so that the safe zone extends to smaller initial viral population sizes as the rate of acquisition increases $\eta \rightarrow f$. Intuitively, large acquisition rates are needed to rapidly degrade the bacterial population and maintain the Lotka-Volterra-like stability.

Conversely, systems with large initial viral population sizes (point C in Fig.~\ref{Fig3}) find themselves in the safe zone only for small $\eta$ values (Fig.~\ref{Fig4}C). In this case, small effective degradation rates keep the bacterial population relatively large and stable compared to the viral population. In this parameter regime, the optimal strategy consists of minimizing spacer acquisition to increase fitness. 

Lastly, for intermediate initial viral population sizes (point D in Fig.~\ref{Fig3}), deterministic dynamics are extinction-free for intermediate $\eta$ values (Fig.~\ref{Fig4}D). Effectively, we recover the situation in Fig.~\ref{Fig4}A, where an intermediate acquisition rate is optimal, albeit for a different reason. In Fig.~\ref{Fig4}A the bacterial population increases its survival probability by stochastically acquiring non-self targeting spacers. However the probability of extinction remains non-zero and the bacteria are never truly safe. For the initial condition parameter regime in Fig.~\ref{Fig4}D, the bacteria will never go extinct with intermediate acquisition rates, simply due to the deterministic Lotka-Volterra-like dynamics \cite{Weissman2018}. By \AW{having different $\eta$ and fitness}, bacteria control the coupled dynamics so as to keep their population size away from extinction. In this case, the acquisition of protective spacers is not essential, similarly to the cases of Fig.~\ref{Fig4}B and C. 

\begin{figure*}
	 \includegraphics[width=\linewidth]{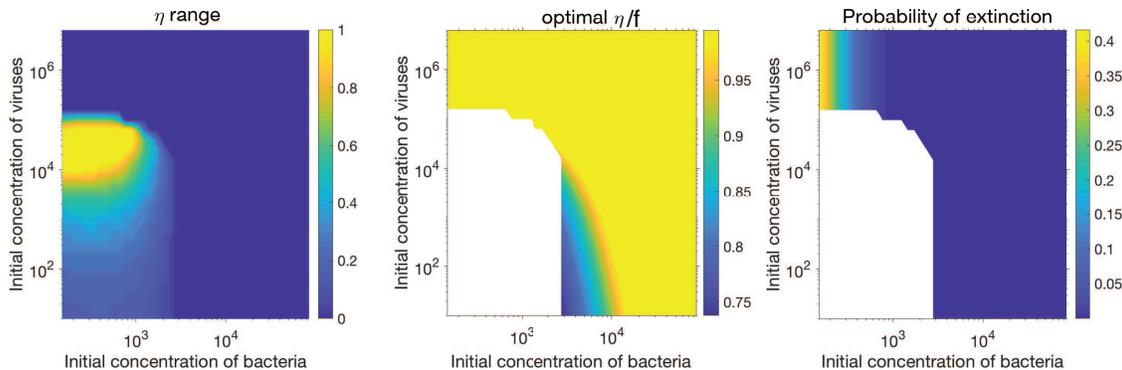}
	{\caption{\label{Fig5}\textbf{Phase diagrams of the optimal survival strategies.} (A) Range of acquisition rates $\eta$ that position the system in the safe zone, as a function of initial virus and bacteria population sizes. The range is defined as the width of the region of $\eta$ values where the extinction probability is strictly zero in Fig.~\ref{Fig4}B-D. The initial conditions of points A-D presented in Fig.~\ref{Fig4} are shown as grey dots. (B) Optimal acquisition rate normalized by the growth rate $f$ for initial population sizes where the safe zone is not available as a function of initial virus and bacteria population sizes. The optimal acquisition rate is defined as the minimum of the probability of extinction in Fig.~\ref{Fig4}A. (C) Probability of extinction associated to the optimal rate. In B and C the white area corresponds to zero extinction probability. Parameters: $K=10^6,  f=0.5\;\mbox{hr}^{-1}$ per cell, $d=0.2\;\mbox{hr}^{-1}$ per phage, $\beta=0.01$ and $\mu=0.8\;\mbox{hr}^{-1}$ per cell. The concentrations are given in terms of mean numbers of individuals.}
}
\end{figure*}

Clearly, \AW{having} acquisition rates such as to fall into the safe zone is the best strategy the bacterial population can adopt, since it guarantees certain survival. From Fig.~\ref{Fig4} we see that the range of $\eta$ values that guarantee survival is larger or smaller, depending on the different initial population sizes. Additionally, in certain cases the safe zone is not accessible.  We can summarize the discussion of Fig.~\ref{Fig4} by two phase diagrams: one that shows the width of the range of acquisition rates that position the system in the safe zone (Fig.~\ref{Fig5}A), and one that shows the optimal acquisition rate for initial population sizes where the safe zone is not available (Fig.~\ref{Fig5}B), along with the extinction probability thus achieved (Fig.~\ref{Fig5}C). 
Moderately large initial viral populations ensure non-extinction dynamics for a broad choice of acquisition rates, and for a wide range of initial bacterial population sizes. The range of $\eta$ values that guarantee safe dynamics shrinks with decreasing initial viral population sizes, and becomes confined to a narrower range of initial bacterial population sizes. This result reflects the compatibility requirement between viral and bacterial populations in the Lotka-Volterra dynamics. The acquisition rate has the role of an effective death term that limits the growth of the bacterial population, in turn limiting viral expansion, and controlling the amplitude of oscillations around the fixed point. If the Lotka-Volterra-like balance is broken, fluctuations become large, and the dynamics leaves the safe zone (Fig.~\ref{Fig5}B). In that case, bacteria rely on stochastically acquiring spacers for survival. For very high initial viral population sizes, the susceptible bacteria find themselves in the kamikaze regime: they are largely out-numbered so the best strategy is to take up as much DNA as they can, not worrying about the drawbacks of auto-immunity. In any case, as we see from Fig.~\ref{Fig3}, in this regime their chance of survival is extremely small. For smaller initial viral population sizes, the danger of auto-immunity kicks in and optimal strategies rely on intermediate $\eta$ values, as described in Fig.~\ref{Fig4}A.

\section{Discussion}

The general question of the efficiency of CRISPR-Cas mediated immunity has already been extensively explored \cite{Shudo2001, Hamilton2008, Koonin2009, Westra2015}. Our approach focuses on how \AW{regulating} the spacer acquisition rate affects the survival probability of bacterial populations. Specifically, we were interested in the dangers of auto-immunity stemming from acquiring self-targeting spacers \cite{Wei2015, Levy2015b}. This danger becomes larger for large acquisition rates and our simple model does show that it is often best for bacterial survival to avoid very large acquisition rates. However, the straightforward argument of wanting to avoid self-acquisition while still targeting phages works only in a limited range of initial phage and bacterial population sizes. For many systems, survival is insured by a deterministic Lotka-Volterra-like dynamics that keeps the two co-evolving population sizes balanced so that neither grows too large. \AW{Having an acquisition rate $\eta$ to an optimal value for survival within this model means having an initial condition that is positioned in this safe zone regime,} where deterministic Lotka-Volterra-like dynamics guarantees convergence to the fixed point, rather than having to rely on a stochastic acquisition event. 

Nevertheless, for certain values of initial population sizes, no value of $\eta$ guarantees certain survival and in these regimes avoiding auto-immunity becomes important. For most initial population sizes in this regime there exists an optimal intermediate value of the acquisition rate that results in a non-zero survival probability, and prevents the whole population being killed by both phages and auto-immunity. Only for very large initial phage populations is the optimal acquisition rate close to the maximum possible value of the maximal growth rate. In this case however, even this optimal choice of $\eta$ fails to substantially ensure survival. However, as previous theory and experiments have shown \cite{Westra2015, Mayer2015b}, in the regime of high phage density and large exposure, constitutive immune strategies such as surface modification mediated immunity are more common than \AW{CRISPR-Cas} mediated immunity. \AW{In our analysis, we have not introduced any other forms of immunity since we wanted to focus only on how CRISPR-Cas self-acquisition affects the survival rate. For this reason we cannot directly compare our results to these experiments~\cite{Westra2015}. However, the poor survival rate of the CRSIPR-Cas system predicted by our model in the regime of high initial density is not inconsistent with experimental results, although the source of the poor survival in the experiment and in the model is likely to be very different. To explore this origin we would need to model receptor resistance explicitly. Our results do however second the theoretical idea~\cite{Mayer2015b} that constitutive immunity is a better strategy for very large phage populations. }

We assumed that the acquisition rate is a parameter the bacterial population can \AW{regulate}, since bacteria make the decision to uptake or not uptake foreign DNA by turning on their adaptive form of immunity \cite{Westra2015, Patterson2016, Hoyland-Kroghsbo2017}. On the other hand we assumed that \AW{modifying $\beta$ -- the probability of up-taking non self-targeting DNA, as opposed to self DNA }-- is much harder \cite{Wei2015, Levy2015b}. In principle, one can imagine that on evolutionary timescales \AW{$\beta$ can be modified} by mutating PAM motifs. However the timescales for this are much longer than the phenotypic modification of \AW{modifying $\eta$}, and this strategy has limited benefits due to the short lengths of PAM motifs. \AW{It would interesting in the future to explore  dynamics that  lead to the optimal value of the acquisition rate $\eta$, and/or allow it to be controlled accurately. However, we do not explore this issue in this work.}

Experimental studies have both reported  the unbiased acquisition of self-targeting spacers from the bacterial genome compared to viral DNA \cite{Wei2015} (the low observed frequency of self spacers being is due to the elimination of self-targeting bacteria), and identified mechanisms for preferential acquisition of viral spacers \cite{Levy2015b} (at stalled replication forks which are more common in phages, and away from Chi sites which are more common in bacteria). Since the acquisition rates in our model are effective, \AW{we can change them and explore both limits. Our results are compatible with both of these scenarios. When increasing the value of $\beta$, we observe that the probability of extinction goes to zero exponentially fast with a characteristic value $\eta_c\ll f$. This means that a small value of $\eta$ ensures almost perfect survival from the infection and a maximum growth rate very similar to the wild type \AW{that is consistent with the results that a functional CRISPR-Cas mechanism do not change much the fitness in competition experiments in absence of the phage}.}

Recent studies have shown that bacteria in both {\em P. aeruginosa} and {\em Serratia} can tune the use of CRISPR-Cas mediated immunity by communicating with other bacteria in the population so that the population level immune defense is coordinated \cite{Patterson2016, Hoyland-Kroghsbo2017}. The coordination is achieved through quorum sensing pathways and leads to an increase in the expression of Cas proteins at high bacterial cell density. Cas proteins are directly involved in spacer acquisition, so \AW{having different Cas protein concentrations is a direct way of having different acquisition rates.}  A theoretical model showed that regulating the CRISPR-Cas  system in a density dependent way, leads to phenotypic bistability in the population \cite{Bonsma-Fisher2018}: CRISPR-Cas immunity is up-regulated at high bacterial density and down-regulated at low bacterial density. The two ecologically different states differ both by their bacteria to phage ratio and by their spacer diversity composition. Quorum sensing introduces an additional feedback mechanism into our model, whose effects need to be studied in more detail. In the monostable regime, this feedback simply renormalizes the acquisition rate. In the bistable regime, it may lead to a more complicated form of the phase diagram but it is unlikely to change the general conclusion about the existence of a safe zone and a regime where the acquisition rate matters. Additionally, it would be interesting to consider a more detailed model that predicts the spacer diversity distribution in the safe zone compared to the acquisition zone. This would provide a means of directly comparing our results to experiments.  


Our results also show that in general up-taking spacers is a good survival strategy. We identified a set of initial population sizes where $\eta$ \AW{can be} strictly equal to zero. Small $\eta$ is the optimal strategy for a small range of initial conditions.  Even if a large fraction of bacteria will die from self-targeted spacers, acquiring DNA is still a good population level survival strategy. \AW{However, if the CRISPR-Cas is able to discriminate self versus non-self, meaning that the probability of acquiring self-targeting spacers is low, then a small value of acquisition rate $\eta$ is enough to ensure almost perfect survival rate of the colony. Thus in this regime, the regulation of acquisition rate is not under selective pressure.}

Experiments show that Cas protein expression, which is an essential part of acquiring new spacers, is the most costly element of maintaining the CRISPR-Cas type II-A system in {\it Streptococcus thermophilus} \cite{Vale2015a}. Cas protein expression reduces fitness, and Cas-deficient mutants achieve higher fitness in competition experiments against WT strains, in contrast to experiments with the type I-F CRISPR-Cas system of {\it P. aeruginosa}, where knocking out a {\it cas} gene  had little effect on its competitive growth rate with respect to the WT strain \cite{Westra2015}. The growth rate reduction in  {\it S. thermophilus}  seem to also depend on phage exposure \cite{Vale2015a}, but no significant fitness costs were identified associated with launching the CRISPR-Cas immune system. Once the CRISPR-Cas system is in place, acquiring spacers does not seem to hold additional costs besides auto-immunity. 

Following recent work \cite{Bonsma-Fisher2018}, our model \AW{stresses}  the importance of population level approaches. It is important that the \AW{susceptible} bacterial population survives and not particular individuals. However, the model we consider is simple and we make a number of approximations. First of all, we treat extinction deterministically: either the bacterial population will go extinct if  it reaches small numbers at the $N_{\rm ext}=10$ boundary, or it will survive. In reality, the dynamics at small population sizes is stochastic and extinction only happens at $N=0$. Even a bacterial population that descends to small numbers can still grow and increase over the arbitrarily chosen deterministic threshold, by means of stochastic dynamics. Conversely, even during the safe deterministic trajectory we can imagine an accumulation of rare acquisition events of self-targeting spacers that will lead the population to extinction.
Considering a full stochastic model will likely make quantitative differences but it is unlikely to change the qualitative conclusions. We also assumed that once there is a first phage targeting acquisition event, the population is sure to survive. However, the lineage carrying the protection could still go extinct due to the stochastic nature of division and death. Taking this effect into account would simply renormalize the probability of foreign spacer acquisition $\beta$, which is an effective parameter of our model, and would not affect our conclusions.

{\bf Acknowledgements.} This work was supported in part by grant ERCCOG n. 724208. We thank Julia Grajek, Zofia Grochulska and Maja Szlenk for helpful discussions during the Simons Semester on Mathematical Biology.

\bibliographystyle{pnas}
\bibliography{serena_crispr,thierry,serena_new}

\begin{thebibliography}{10}

\bibitem{Medzhitov1997}
Medzhitov R, Janeway CAJ
\newblock (1997) {Innate immunity: Minireview the virtues of a nonclonal system
  of recognition}.
\newblock \emph{Cell} 91:295--298.

\bibitem{Boehm2011}
Boehm T
\newblock (2011) {Design principles of adaptive immune systems}.
\newblock \emph{Nature Reviews Immunology} 11:307--317.

\bibitem{Spoel2012}
Spoel SH, Dong X
\newblock (2012) {How do plants achieve immunity? Defence without specialized
  immune cells}.
\newblock \emph{Nature Reviews Immunology} 12:89--100.

\bibitem{Murphy2007}
Murphy K, Travers P, Walport M
\newblock (2007) \emph{{Janeway's Immunology}}
\newblock (Garland Science), 7 edition edition.

\bibitem{Mayer2015b}
Mayer A, Mora T, Rivoire O, Walczak AM
\newblock (2015) {Diversity of immune strategies explained by adaptation to
  pathogen statistics}.
\newblock \emph{Proceedings of the National Academy of Sciences} 113.

\bibitem{Levins1968}
Levins R
\newblock (1968) \emph{{Evolution in changing environments: some theoretical
  explorations}}
\newblock p 120.

\bibitem{Seger1987}
Seger J, Brockmann HJ
\newblock (1987) {What is bet-hedging?}
\newblock \emph{Oxford Surveys in Evolutionary Biology} 4:182--211.

\bibitem{Donaldson-Matasci2008}
Donaldson-Matasci M, Lachmann M, Bergstrom C
\newblock (2008) {Phenotypic diversity as an adaptation to environmental
  uncertainty}.
\newblock \emph{Evolutionary Ecology Research} 10:493--515.

\bibitem{Mayer2017}
Mayer A, Mora T, Rivoire O, Walczak AM
\newblock (2017) {Transitions in optimal adaptive strategies for populations in
  fluctuating environments}.
\newblock \emph{Physical Review E} 96:1--16.

\bibitem{Abedon2012}
Abedon ST
\newblock (2012) {Bacterial `immunity' against bacteriophages}.
\newblock \emph{Bacteriophage} 2:50--54.

\bibitem{Barrangou2014}
Barrangou R, Marraffini LA
\newblock (2014) {CRISPR-cas systems: Prokaryotes upgrade to adaptive
  immunity}.
\newblock \emph{Molecular Cell} 54:234--244.

\bibitem{Labrie2010}
Labrie SJ, Samson JE, Moineau S
\newblock (2010) {Bacteriophage resistance mechanisms}.
\newblock \emph{Nature Reviews Microbiology} 8:317--327.

\bibitem{Bonsma-Fisher2018}
Bonsma-Fisher M, Soutiere D, Goyal S
\newblock (2018) {How adaptive immunity constrains composition and fate of
  large bacterial populations}.
\newblock \emph{Proceedings of the National Academy of Sciences}.

\bibitem{Heidelberg2009}
Heidelberg JF, Nelson WC, Schoenfeld T, Bhaya D
\newblock (2009) {Germ warfare in a microbial mat community: CRISPRs provide
  insights into the co-evolution of host and viral genomes}.
\newblock \emph{PLoS ONE} 4.

\bibitem{Andersson2008}
Andersson AF, Banfield JF
\newblock (2008) {Virus population dynamics and acquired virus resistance in
  natural microbial communities}.
\newblock \emph{Science} 320:1047--1050.

\bibitem{Tyson2008}
Tyson GW, Banfield JF
\newblock (2008) {Rapidly evolving CRISPRs implicated in acquired resistance of
  microorganisms to viruses}.
\newblock \emph{Environmental Microbiology} 10:200--207.

\bibitem{Paez-Espino2015}
Paez-Espino D, {et~al.}
\newblock (2015) {CRISPR immunity drives rapid phage genome evolution in
  streptococcus thermophilus}.
\newblock \emph{mBio} 6:1--9.

\bibitem{Mojica2009}
Mojica FJ, D{\'{i}}ez-Villase{\~{n}}or C, Garc{\'{i}}a-Mart{\'{i}}nez J,
  Almendros C
\newblock (2009) {Short motif sequences determine the targets of the
  prokaryotic CRISPR defence system}.
\newblock \emph{Microbiology} 155:733--740.

\bibitem{Marraffini2010a}
Marraffini LA, Sontheimer EJ
\newblock (2010) {Self versus non-self discrimination during CRISPR
  RNA-directed immunity}.
\newblock \emph{Nature} 463:568--571.

\bibitem{Levy2015b}
Levy A, {et~al.}
\newblock (2015) {CRISPR adaptation biases explain preference for acquisition
  of foreign DNA}.
\newblock \emph{Nature} 520:505--510.

\bibitem{Wei2015}
Wei Y, Terns RM, Terns MP
\newblock (2015) {Cas9 function and host genome sampling in type II-A
  CRISPR--cas adaptation}.
\newblock \emph{Genes and Development} 29:356--361.

\bibitem{Westra2015}
Westra ER, {et~al.}
\newblock (2015) {Parasite exposure drives selective evolution of constitutive
  versus inducible defense}.
\newblock \emph{Current Biology} 25:1043--1049.

\bibitem{Stern2010}
Stern A, Keren L, Wurtzel O, Amitai G, Sorek R
\newblock (2010) {Self-targeting by CRIPR: gene regulation or autoimmunity?}
\newblock \emph{Trends Genet.} 26:335--340.

\bibitem{Vercoe2013}
Vercoe RB, {et~al.}
\newblock (2013) {Cytotoxic Chromosomal Targeting by CRISPR/Cas Systems Can
  Reshape Bacterial Genomes and Expel or Remodel Pathogenicity Islands}.
\newblock \emph{PLoS Genetics} 9.

\bibitem{Edgar2010}
Edgar R, Qimron U
\newblock (2010) {The Escherichia coli CRISPR system protects from $\lambda$
  lysogenization, lysogens, and prophage induction}.
\newblock \emph{Journal of Bacteriology} 192:6291--6294.

\bibitem{Berezovskaya2014}
Berezovskaya FS, Wolf YI, Koonin EV, Karev GP
\newblock (2014) {Pseudo-chaotic oscillations in CRISPR-virus coevolution
  predicted by bifurcation analysis}.
\newblock \emph{Biology Direct} 9:1--17.

\bibitem{Childs2012}
Childs LM, {et~al.}
\newblock (2012) {Multi-scale model of CRISPR-induced coevolutionary dynamics:
  diversification at the interface of Lamarck and Darwin}.
\newblock \emph{Evolution} pp 2014--2029.

\bibitem{Childs2014}
Childs LM, England WE, Young MJ, Weitz JS, Whitaker RJ
\newblock (2014) {CRISPR-induced distributed immunity in microbial
  populations}.
\newblock \emph{PLoS ONE} 9:1--12.

\bibitem{Han2017}
Han P, Deem MW
\newblock (2017) {Non-classical phase diagram for virus bacterial coevolution
  mediated by clustered regularly interspaced short palindromic repeats}.
\newblock \emph{Journal of the Royal Society Interface} 14.

\bibitem{He2010}
He J, Deem MW
\newblock (2010) {Heterogeneous Diversity of Spacers within CRISPR (Clustered
  Regularly Interspaced Short Palindromic Repeats)}.
\newblock \emph{Physical Review Letters} 105:128102.

\bibitem{Haerter2011}
Haerter JO, Trusina A, Sneppen K
\newblock (2011) {Targeted Bacterial Immunity Buffers Phage Diversity}.
\newblock \emph{Journal of Virology} 85:10554--10560.

\bibitem{Weinberger2012a}
Weinberger AD, Wolf YI, Lobkovsky AE, Gilmore MS, Koonin EV
\newblock (2012) {Viral diversity threshold for adaptive immunity in
  prokaryotes}.
\newblock \emph{mBio} 3:1--10.

\bibitem{Bradde2017}
Bradde S, Vucelja M, Te{\c s}ileanu T, Balasubramanian V
\newblock (2017) {Dynamics of adaptive immunity against phage in bacterial
  populations}.
\newblock \emph{PLoS Computational Biology} 13:1--16.

\bibitem{Koonin2009}
Koonin EV, Wolf YI
\newblock (2009) {Is evolution Darwinian or/and Lamarckian?}
\newblock \emph{Biol. Direct} 4:42.

\bibitem{Haerter2012}
Haerter JO, Sneppen K
\newblock (2012) {Spatial structure and Lamarckian adaptation explain extreme
  genetic diversity at CRISPR Locus}.
\newblock \emph{mBio} 3:1--6.

\bibitem{Iranzo2013}
Iranzo J, Lobkovsky AE, Wolf YI, Koonin EV
\newblock (2013) {Evolutionary dynamics of the prokaryotic adaptive immunity
  system CRISPR-Cas in an explicit ecological context}.
\newblock \emph{Journal of Bacteriology} 195:3834--3844.

\bibitem{Levin2013}
Levin BR, Moineau S, Bushman M, Barrangou R
\newblock (2013) {The Population and Evolutionary Dynamics of Phage and
  Bacteria with CRISPR-Mediated Immunity}.
\newblock \emph{PLoS Genetics} 9.

\bibitem{PleskaNatEE2018}
Ple{\v{s}}ka M, Lang M, Refardt D, Levin BR, Guet CC
\newblock (2018) {Phage–host population dynamics promotes prophage
  acquisition in bacteria with innate immunity}.
\newblock \emph{Nature Ecology and Evolution}.

\bibitem{Levin2010}
Levin BR
\newblock (2010) {Nasty viruses, costly plasmids, population dynamics, and the
  conditions for establishing and maintaining CRISPR-mediated adaptive immunity
  in bacteria}.
\newblock \emph{PLoS Genetics} 6:1--12.

\bibitem{Vale2015a}
Vale PF, {et~al.}
\newblock (2015) {Costs of CRISPR-Cas-mediated resistance in Streptococcus
  thermophilus}.
\newblock \emph{Proceedings of the Royal Society B: Biological Sciences} 282.

\bibitem{Pleska2016}
Pleska M, {et~al.}
\newblock (2016) {Bacterial Autoimmunity Due to a Restriction- Modification
  System}.
\newblock \emph{Current Biology} 26:404--409.

\bibitem{Perelson1997}
Perelson A, Weisbuch G
\newblock (1997) {Immunology for physicists}.
\newblock \emph{Reviews of Modern Physics} 69:1219--1268.

\bibitem{Datsenko2012}
Datsenko KA, {et~al.}
\newblock (2012) {Molecular memory of prior infections activates the CRISPR/Cas
  adaptive bacterial immunity system}.
\newblock \emph{Nature Communications} 3:945--947.

\bibitem{Patterson2016}
Patterson AG, {et~al.}
\newblock (2016) {Quorum Sensing Controls Adaptive Immunity through the
  Regulation of Multiple CRISPR-Cas Systems}.
\newblock \emph{Molecular Cell} 64:1102--1108.

\bibitem{Hoyland-Kroghsbo2017}
H{\o}yland-Kroghsbo NM, {et~al.}
\newblock (2017) {Quorum sensing controls the Pseudomonas aeruginosa CRISPR-Cas
  adaptive immune system}.
\newblock \emph{Proceedings of the National Academy of Sciences} 114:131--135.

\bibitem{Lotka20}
Lotka AJ
\newblock (1920) {Analytical note on certain rhythmic relations in organic
  systems}.
\newblock \emph{Proc. Nat. Acad} 6:410--415.

\bibitem{Volterra56}
Volterra V
\newblock (1956) \emph{{Opere mathematiche. Memorie e Note.}}
\newblock (Accademia Nazionale dei Lincei, Rome), pp xxxiii + 604.

\bibitem{Levin77}
Levin BR, {et~al.}
\newblock (2015) {Resource-Limited Growth, Competition, and Predation: A Model
  and Experimental Studies with Bacteria and Bacteriophage}.
\newblock \emph{American Naturalist} 111:3--24.

\bibitem{Weitz:2015aa}
Weitz J
\newblock (2015) Quantitative viral ecology : dynamics of viruses and their
  microbial hosts.

\bibitem{Shudo2001}
Shudo E, Iwasa Y
\newblock (2001) {Inducible defense against pathogens and parasites: Optimal
  choice among multiple options}.
\newblock \emph{Journal of Theoretical Biology} 209:233--247.

\bibitem{Hamilton2008}
Hamilton R, Siva-Jothy M, Boots M
\newblock (2008) {Two arms are better than one: Parasite variation leads to
  combined inducible and constitutive innate immune responses}.
\newblock \emph{Proceedings of the Royal Society B: Biological Sciences}
  275:937--945.

\bibitem{Heler2017}
Heler R, {et~al.}
\newblock (2017) {Mutations in Cas9 Enhance the Rate of Acquisition of Viral
  Spacer Sequences during the CRISPR-Cas Immune Response}.
\newblock \emph{Molecular Cell} 65:168--175.

\bibitem{Weissman2018}
Weissman JL, {et~al.}
\newblock (2018) {Immune loss as a driver of coexistence during host-phage
  coevolution}.
\newblock \emph{ISME Journal} 12:585--597.

\end{thebibliography}

\end{document}